\pdfoutput=1
\documentclass[12pt]{article}

\usepackage{graphicx}

\input{epsf}

\advance \topmargin by -\headheight
\advance \topmargin by -\headsep

\evensidemargin \oddsidemargin
\begin{document}

\topmargin -.6in

\def\rh{{\hat \rho}}
\def\alie{{\hat{\cal G}}}
\newcommand{\sect}[1]{\setcounter{equation}{0}\section{#1}}
\renewcommand{\theequation}{\thesection.\arabic{equation}}

\def\rf#1{(\ref{eq:#1})}
\def\lab#1{\label{eq:#1}}
\def\nonu{\nonumber}
\def\br{\begin{eqnarray}}
\def\er{\end{eqnarray}}
\def\be{\begin{equation}}
\def\ee{\end{equation}}
\def\eq{\!\!\!\! &=& \!\!\!\! }
\def\foot#1{\footnotemark\footnotetext{#1}}
\def\lb{\lbrack}
\def\rb{\rbrack}
\def\llangle{\left\langle}
\def\rrangle{\right\rangle}
\def\blangle{\Bigl\langle}
\def\brangle{\Bigr\rangle}
\def\llbrack{\left\lbrack}
\def\rrbrack{\right\rbrack}
\def\lcurl{\left\{}
\def\rcurl{\right\}}
\def\({\left(}
\def\){\right)}
\newcommand{\nit}{\noindent}
\newcommand{\ct}[1]{\cite{#1}}
\newcommand{\bi}[1]{\bibitem{#1}}
\def\lskip{\vskip\baselineskip\vskip-\parskip\noindent}
\relax

\def\tr{\mathop{\rm tr}}
\def\Tr{\mathop{\rm Tr}}
\def\v{\vert}
\def\bv{\bigm\vert}
\def\Bgv{\;\Bigg\vert}
\def\bgv{\bigg\vert}
\newcommand\partder[2]{{{\partial {#1}}\over{\partial {#2}}}}
\newcommand\funcder[2]{{{\delta {#1}}\over{\delta {#2}}}}
\newcommand\Bil[2]{\Bigl\langle {#1} \Bigg\vert {#2} \Bigr\rangle}  
\newcommand\bil[2]{\left\langle {#1} \bigg\vert {#2} \right\rangle} 
\newcommand\me[2]{\left\langle {#1}\bv {#2} \right\rangle} 
\newcommand\sbr[2]{\left\lbrack\,{#1}\, ,\,{#2}\,\right\rbrack}
\newcommand\pbr[2]{\{\,{#1}\, ,\,{#2}\,\}}
\newcommand\pbbr[2]{\lcurl\,{#1}\, ,\,{#2}\,\rcurl}

\def\ket#1{\mid {#1} \rangle}
\def\bra#1{\langle {#1} \mid}
\newcommand{\braket}[2]{\langle {#1} \mid {#2}\rangle}
%
\def\a{\alpha}
\def\at{{\tilde A}^R}
\def\atc{{\tilde {\cal A}}^R}
\def\atcm#1{{\tilde {\cal A}}^{(R,#1)}}
\def\b{\beta}
\def\dc{{\cal D}}
\def\d{\delta}
\def\D{\Delta}
\def\eps{\epsilon}
\def\vareps{\varepsilon}
\def\g{\gamma}
\def\G{\Gamma}
\def\grad{\nabla}
\def\h{{1\over 2}}
\def\l{\lambda}
\def\L{\Lambda}
\def\m{\mu}
\def\n{\nu}
\def\o{\over}
\def\om{\omega}
\def\O{\Omega}
\def\p{\phi}
\def\P{\Phi}
\def\pa{\partial}
\def\pr{\prime}
\def\pt{{\tilde \Phi}}
\def\qs{Q_{\bf s}}
\def\ra{\rightarrow}
\def\s{\sigma}
\def\S{\Sigma}
\def\t{\tau}
\def\th{\theta}
\def\Th{\Theta}
\def\tpp{\Theta_{+}}
\def\tmm{\Theta_{-}}
\def\tpg{\Theta_{+}^{>}}
\def\tms{\Theta_{-}^{<}}
\def\tp0{\Theta_{+}^{(0)}}
\def\tm0{\Theta_{-}^{(0)}}
\def\ti{\tilde}
\def\wti{\widetilde}
\def\jc{J^C}
\def\bj{{\bar J}}
\def\sj{{\jmath}}
\def\bsj{{\bar \jmath}}
\def\bp{{\bar \p}}
\def\vp{\varphi}
\def\ve{\varepsilon}
\def\vt{{\tilde \varphi}}
\def\faa{Fa\'a di Bruno~}
\def\ca{{\cal A}}
\def\cb{{\cal B}}
\def\ce{{\cal E}}
\def\cg{{\cal G}}
\def\cgh{{\hat {\cal G}}}
\def\ch{{\cal H}}
\def\chh{{\hat {\cal H}}}
\def\cl{{\cal L}}
\def\cm{{\cal M}}
\def\cn{{\cal N}}
\def\u2{\mid u\mid^2}
\newcommand\sumi[1]{\sum_{#1}^{\infty}}   
\newcommand\fourmat[4]{\left(\begin{array}{cc}  
{#1} & {#2} \\ {#3} & {#4} \end{array} \right)}

%
\def\lie{{\cal G}}
\def\kmlie{{\hat{\cal G}}}
\def\dlie{{\cal G}^{\ast}}
\def\elie{{\widetilde \lie}}
\def\edlie{{\elie}^{\ast}}
\def\hlie{{\cal H}}
\def\flie{{\cal F}}
\def\wlie{{\widetilde \lie}}
\def\f#1#2#3 {f^{#1#2}_{#3}}
\def\winf{{\sf w_\infty}}
\def\win1{{\sf w_{1+\infty}}}
\def\hwinf{{\sf {\hat w}_{\infty}}}
\def\Winf{{\sf W_\infty}}
\def\Win1{{\sf W_{1+\infty}}}
\def\hWinf{{\sf {\hat W}_{\infty}}}
\def\Rm#1#2{r(\vec{#1},\vec{#2})}          
\def\OR#1{{\cal O}(R_{#1})}           
\def\ORti{{\cal O}({\widetilde R})}           
\def\AdR#1{Ad_{R_{#1}}}              
\def\dAdR#1{Ad_{R_{#1}^{\ast}}}      
\def\adR#1{ad_{R_{#1}^{\ast}}}       
\def\KP{${\rm \, KP\,}$}                 
\def\KPl{${\rm \,KP}_{\ell}\,$}         
\def\KPo{${\rm \,KP}_{\ell = 0}\,$}         
\def\mKPa{${\rm \,KP}_{\ell = 1}\,$}    
\def\mKPb{${\rm \,KP}_{\ell = 2}\,$}    
%
\def\rlx{\relax\leavevmode}
\def\inbar{\vrule height1.5ex width.4pt depth0pt}
\def\IZ{\rlx\hbox{\sf Z\kern-.4em Z}}
\def\IR{\rlx\hbox{\rm I\kern-.18em R}}
\def\IC{\rlx\hbox{\,$\inbar\kern-.3em{\rm C}$}}
\def\IN{\rlx\hbox{\rm I\kern-.18em N}}
\def\IO{\rlx\hbox{\,$\inbar\kern-.3em{\rm O}$}}
\def\IP{\rlx\hbox{\rm I\kern-.18em P}}
\def\IQ{\rlx\hbox{\,$\inbar\kern-.3em{\rm Q}$}}
\def\IF{\rlx\hbox{\rm I\kern-.18em F}}
\def\IG{\rlx\hbox{\,$\inbar\kern-.3em{\rm G}$}}
\def\IH{\rlx\hbox{\rm I\kern-.18em H}}
\def\II{\rlx\hbox{\rm I\kern-.18em I}}
\def\IK{\rlx\hbox{\rm I\kern-.18em K}}
\def\IL{\rlx\hbox{\rm I\kern-.18em L}}
\def\one{\hbox{{1}\kern-.25em\hbox{l}}}
\def\0#1{\relax\ifmmode\mathaccent"7017{#1}%
B        \else\accent23#1\relax\fi}
\def\omz{\0 \omega}
%
\def\ltimes{\mathrel{\vrule height1ex}\joinrel\mathrel\times}
\def\rtimes{\mathrel\times\joinrel\mathrel{\vrule height1ex}}
%
\def\mark{\noindent{\bf Remark.}\quad}
\def\prop{\noindent{\bf Proposition.}\quad}
\def\theor{\noindent{\bf Theorem.}\quad}
\def\name{\noindent{\bf Definition.}\quad}
\def\exam{\noindent{\bf Example.}\quad}
\def\proof{\noindent{\bf Proof.}\quad}
%
%
\def\PRL#1#2#3{{\sl Phys. Rev. Lett.} {\bf#1} (#2) #3}
\def\NPB#1#2#3{{\sl Nucl. Phys.} {\bf B#1} (#2) #3}
\def\NPBFS#1#2#3#4{{\sl Nucl. Phys.} {\bf B#2} [FS#1] (#3) #4}
\def\CMP#1#2#3{{\sl Commun. Math. Phys.} {\bf #1} (#2) #3}
\def\PRD#1#2#3{{\sl Phys. Rev.} {\bf D#1} (#2) #3}
\def\PLA#1#2#3{{\sl Phys. Lett.} {\bf #1A} (#2) #3}
\def\PLB#1#2#3{{\sl Phys. Lett.} {\bf #1B} (#2) #3}
\def\JMP#1#2#3{{\sl J. Math. Phys.} {\bf #1} (#2) #3}
\def\PTP#1#2#3{{\sl Prog. Theor. Phys.} {\bf #1} (#2) #3}
\def\SPTP#1#2#3{{\sl Suppl. Prog. Theor. Phys.} {\bf #1} (#2) #3}
\def\AoP#1#2#3{{\sl Ann. of Phys.} {\bf #1} (#2) #3}
\def\PNAS#1#2#3{{\sl Proc. Natl. Acad. Sci. USA} {\bf #1} (#2) #3}
\def\RMP#1#2#3{{\sl Rev. Mod. Phys.} {\bf #1} (#2) #3}
\def\PR#1#2#3{{\sl Phys. Reports} {\bf #1} (#2) #3}
\def\AoM#1#2#3{{\sl Ann. of Math.} {\bf #1} (#2) #3}
\def\UMN#1#2#3{{\sl Usp. Mat. Nauk} {\bf #1} (#2) #3}
\def\FAP#1#2#3{{\sl Funkt. Anal. Prilozheniya} {\bf #1} (#2) #3}
\def\FAaIA#1#2#3{{\sl Functional Analysis and Its Application} {\bf #1} (#2)
#3}
\def\BAMS#1#2#3{{\sl Bull. Am. Math. Soc.} {\bf #1} (#2) #3}
\def\TAMS#1#2#3{{\sl Trans. Am. Math. Soc.} {\bf #1} (#2) #3}
\def\InvM#1#2#3{{\sl Invent. Math.} {\bf #1} (#2) #3}
\def\LMP#1#2#3{{\sl Letters in Math. Phys.} {\bf #1} (#2) #3}
\def\IJMPA#1#2#3{{\sl Int. J. Mod. Phys.} {\bf A#1} (#2) #3}
\def\AdM#1#2#3{{\sl Advances in Math.} {\bf #1} (#2) #3}
\def\RMaP#1#2#3{{\sl Reports on Math. Phys.} {\bf #1} (#2) #3}
\def\IJM#1#2#3{{\sl Ill. J. Math.} {\bf #1} (#2) #3}
\def\APP#1#2#3{{\sl Acta Phys. Polon.} {\bf #1} (#2) #3}
\def\TMP#1#2#3{{\sl Theor. Mat. Phys.} {\bf #1} (#2) #3}
\def\JPA#1#2#3{{\sl J. Physics} {\bf A#1} (#2) #3}
\def\JSM#1#2#3{{\sl J. Soviet Math.} {\bf #1} (#2) #3}
\def\MPLA#1#2#3{{\sl Mod. Phys. Lett.} {\bf A#1} (#2) #3}
\def\JETP#1#2#3{{\sl Sov. Phys. JETP} {\bf #1} (#2) #3}
\def\JETPL#1#2#3{{\sl  Sov. Phys. JETP Lett.} {\bf #1} (#2) #3}
\def\PHSA#1#2#3{{\sl Physica} {\bf A#1} (#2) #3}
\def\PHSD#1#2#3{{\sl Physica} {\bf D#1} (#2) #3}
\def\PJA#1#2#3{{\sl Proc. Japan. Acad} {\bf #1A} (#2) #3}
\def\JPSJ#1#2#3{{\sl J. Phys. Soc. Japan} {\bf #1} (#2) #3}

\begin{titlepage}
\vspace*{-1cm}

\vskip 3cm

\vspace{.2in}
\begin{center}
{\large\bf Exact self-duality in a modified Skyrme model }
\end{center}

\vspace{.5cm}

\begin{center}
L. A. Ferreira

\vspace{.3 in}
\small

\par \vskip .2in \noindent
Instituto de F\'\i sica de S\~ao Carlos; IFSC/USP;\\
Universidade de S\~ao Paulo, USP  \\ 
Caixa Postal 369, CEP 13560-970, S\~ao Carlos-SP, Brazil\\

\normalsize
\end{center}

\vspace{.5in}

\begin{abstract}

We propose a modification of the Skyrme model that supports a self-dual sector possessing exact non-trivial finite energy solutions. The action of such a theory possesses the usual quadratic and quartic terms in field derivatives, but the couplings of the components of the Maurer-Cartan form of the Skyrme model  is made by a non-constant symmetric matrix,  instead of the usual Killing form of the $SU(2)$ Lie algebra. The introduction of such a matrix make the self-duality equations conformally invariant in three space dimensions, even though it may break the global internal symmetries of the original Skyrme model. For the case where that matrix is proportional to the identity we show that the theory possesses exact self-dual Skyrmions of unity topological charges.

\end{abstract} 
\end{titlepage}

\section{Introduction}
\label{sec:intro}
\setcounter{equation}{0}

Self-dual sectors of field theories in various dimensions are characterized by first order differential equations such that their solutions also solve  the second order Euler-Lagrange equations. The  beauty of finding classical solutions with one integration less is not due to dynamical conservation laws, but to the existence of a topological charge possessing an integral representation. Any infinitesimal smooth variation of that functional, representing the topological charge, vanishes for any field configuration, irrespective of it being a classical solution or not. Therefore, such a variational procedure leads to an identity that works like  an Euler-Lagrange equation. Such identity  together with the first order self-duality equations imply the true second order Euler-Lagrange equations of the theory.  

Such an observation  has been used recently to invert the process, and construct theories with a self-dual sector \cite{selfdual,ua845,shnir}. Starting with a given integral representation of a topological charge, one looks for ways of splitting the integrand as the product of two pieces. Squaring and adding those two pieces one obtains the static energy  or Euclidean action of a theory that by construction has an exact self-dual sector. The usual BPS bound on the static energy or Euclidean action, comes as  byproduct of the procedure.   

In this paper we apply such a technique to the barionic topological charge of the usual Skyrme model \cite{skyrme,mantonbook}. We obtain a generalization of the Skyrme  model that possesses, in addition to the three $SU(2)$ fields, a symmetric  three dimensional matrix $h_{ab}$, which is introduced through the process of splitting the integrand of the topological charge functional.  The action of such a model has the usual quadratic and quartic terms in derivatives of the fields, but the group indices are contracted not with  the  Killing form of the $SU(2)$ algebra, but with that symmetric matrix.  It is very intriguing that such a generalization of the Skyrme model possesses an exact self-dual sector with non-trivial finite energy solutions.    In addition to that, the introduction of $h_{ab}$  makes the self-dual sector to be conformally invariant in three space dimensions. On the other hand, depending on the form of that matrix, the usual global symmetries of the Skyrme model are broken down to a given subgroup.  For the case where $h_{ab}$ is the identity matrix times a scalar field, we show that the self-dual sector possesses  Skyrmion solutions of topological charges equal to $\pm 1$. Those solutions are very similar to the usual spherically symmetric Skyrmion, but the radial profile function is exact, and the energy density decays as $1/r^6$, as the radial distance $r$ goes to infinity. 

The nature of the matrix $h_{ab}$, that makes the self-dual sector and its conformal symmetry  possible, is still to be understood. Since it changes the coupling of the components of the Maurer-Cartan form of the original Skyrme model, and since its entries depend upon the space-time coordinates, one can perhaps interpret $h_{ab}$ as some low energy expectation values of  fields of a more fundamental theory in higher energies, that play the role of coupling constants in the effective theory. In this sense, one should consider our model as a low energy effective theory. On the other hand, at the classical level one can perhaps interpret $h_{ab}$ as new independent fields, despite the fact that our construction did not require the introduction of a kinetic term for them. In fact, we show  that if one really treats $h_{ab}$ as independent fields, then their corresponding Euler-Lagrange equations are also solved by the  solutions of the self-dual sector. That is an intriguing fact because the technique of splitting the topological charge density is designed to provide self-dual solutions for the Euler-Lagrange equations associated to the $SU(2)$ Skyrme fields only. 
Our generalization of the Skyrme model is on the same lines as that proposed in \cite{shnir}, that constructed a Skyrme type model with target space $S^3$, with a self-dual sector and an infinity of exact self-dual Skyrmion solutions. There,  the extra field can also be interpreted as lower energy couplings depending upon the space-time coordinates.    

It is worth mentioning that the so-called BPS Skyrme model \cite{adam1,adam2} can also be obtained by a splitting of the density of the barionic topological charge of the Skyrme model. That splitting however does not introduce extra fields, and it leads to an  action  made of two terms, one sextic in derivatives of the fields and the other being a potential. More recently, further self-dual Skyrme models were constructed with similar techniques \cite{adam3,adam4}. It would be interesting to investigate if our methods connect with  another approach to self-dual sectors of Skyrme models \cite{sutcliffe},   that couples the theory to an infinite tower of mesons fields. 

The paper is organized as follows. In section \ref{sec:model} we introduce the model and present its main properties. The  construction of the self-dual sector through the technique of splitting the topological charge is given in section \ref{sec:selfdual}. The conformal symmetry of the self-dual sector is presented in section  \ref{sec:conformal}. The exact self-dual spherically symmetric Skyrmion solutions of unity topological charges are constructed in section \ref{sec:su2solutions}.  In section \ref{sec:conclusions} we conclude with some comments on possible consequences and application of our model. 

\section{The model}
\label{sec:model}
\setcounter{equation}{0}

We shall consider a model, on a four dimensional Minkowski space-time, defined by the action 
\be
S= \int d^4x\left[ \frac{m_0^2}{2}\, h_{ab}\,R^a_{\mu}\,R^{b\,,\, \mu}-\frac{1}{4\,e_0^2}\, h^{-1}_{ab}\,H^a_{\mu\nu}\,H^{b\,,\,\mu\nu}\right]
\lab{model}
\ee
where $m_0$ and $e_0$ are coupling constants, and  $R^a_{\mu}$, $\mu=0,1,2,3$, are the components of the $SU(2)$ Maurer-Cartan form given by
\be
R_{\mu} \equiv i\,\partial_{\mu}U\,U^{-1}\equiv R^a_{\mu}\,T_a \qquad\qquad\qquad 
\lab{rdef}
\ee
with $U$ being an element of the group $SU(2)$, and $T_a$, $a=1,2,3$,  being the generators of the corresponding Lie algebra
\be
\sbr{T_a}{T_b}=i\,\ve_{abc}\,T_c 
\ee
In addition we have that
\be
H^a_{\mu\nu}\equiv \partial_{\mu} R^a_{\nu}-\partial_{\nu} R^a_{\mu}
\ee
and $h_{ab}$ is a symmetric invertible matrix, that perhaps can be considered as new independent fields, or even a functional of the $SU(2)$ fields. The nature of such a matrix is still an open problem and we shall treat it more in detail as we discuss the properties of the model.   We shall use the trace form of the $SU(2)$ Lie algebra 
\be
{\rm Tr}\(T_a\,T_b\)=\kappa\, \delta_{ab}
\ee
where $\kappa$ is a constant that depends upon the representation. In fact, $\kappa=1/2$, for the spinor representation and $\kappa=2$, for the triplet representation. Then one can write 
\be
R^a_{\mu}=\frac{i}{\kappa}\, {\rm Tr}\(\partial_{\mu}U\,U^{-1}\,T_a\)
\lab{rdef2}
\ee
Note that since $U$ is a unitary matrix and $T_a$ are hermitian matrices, it follows that $R^a_{\mu}$ are real quantities. In addition,  $R_{\mu}$ satisfies the Maurer-Cartan equation
\be
\partial_{\mu}R_{\nu}-\partial_{\nu}R_{\mu}+i\,\sbr{R_{\mu}}{R_{\nu}}=0
\lab{maurercartan} 
\ee
and so one can write 
\be
H^a_{\mu\nu}= -\frac{i}{\kappa}\, {\rm Tr}\(\sbr{R_{\mu}}{R_{\nu}}\,T_a\)
\lab{hdef2}
\ee
Therefore, in the case where  $h_{ab}$ is the unit matrix, the theory \rf{model} becomes the original Skyrme model \cite{skyrme,mantonbook}. Note that $R_{\mu}^a$ is invariant under the global right transformations
\be
U\rightarrow U\, g_R \qquad\qquad\qquad \mbox{\rm we have that} \qquad\qquad\qquad R_{\mu}^a\rightarrow R_{\mu}^a
\lab{righttransf}
\ee
with $g_R\in SU(2)$. On the other hand, under the global left transformations, we have
\be
U\rightarrow g_L\, U \qquad\qquad\qquad \mbox{\rm we have that} \qquad\qquad\qquad R_{\mu}^a\rightarrow R_{\mu}^b\, d_{ba}\(g_L^{-1}\)
\lab{lefttransf}
\ee
where $d$ is the adjoint (triplet)  matrix representation of  $SU(2)$, i.e. $g\, T_a\,g^{-1}=T_b\, d_{ba}\(g\)$. The adjoint is a real and unitary representation, and so $d$ is an orthogonal matrix, i.e. $d^{T}=d^{-1}$. 
Therefore, if the matrix $h_{ab}$ is invariant under the adjoint, i.e. $d\, h\,d^T=h$, the theory \rf{model} is invariant under the  group $SU(2)_L\otimes SU(2)_R$, of the transformations \rf{righttransf} and \rf{lefttransf}. However, if it is invariant under a given subgroup ${\cal H}$, the symmetry is broken down to ${\cal H}\otimes SU(2)_R$. So, the first consequence of the introduction of the quantity $h_{ab}$, is the breakdown of the global symmetries of the original Skyrme model. On the other hand, the same quantity $h_{ab}$ enlarges the space-time symmetries of the theory \rf{model}. As we will see below, when $h_{ab}$ is non-constant, the static sector of the theory \rf{model} is conformally invariant in three space dimensions, and that favors the existence of a self-dual sector.   

\section{The self-dual sector}
\label{sec:selfdual}
\setcounter{equation}{0}

For finite energy solutions, the $SU(2)$ fields have to go to a constant at spatial infinity, and so, as long as topological arguments are concerned, one can compactify the three dimensional space $\IR^3$ into the three sphere $S^3$, and the field configurations can be classified by the winding number of the map $S^3\rightarrow SU(2)\equiv S^3$, with the following integral representation 
 \be
 Q= \frac{i}{24\,\pi^2}\int d^3x\; \ve_{ijk}\,{\rm Tr}\(R_i\,R_j\,R_k\)
 \lab{topcharge}
 \ee
with $R_i$ given in  \rf{rdef}. We now follow the reasonings on self-duality of \cite{selfdual,ua845,shnir} to construct a self-dual sector for the theory \rf{model}.  We write \rf{topcharge} as
\br
Q&=& \frac{i}{48\,\pi^2}\int d^3x\; \ve_{ijk}\,{\rm Tr}\(R_i\,\sbr{R_j}{R_k}\)
 =-\frac{1}{48\,\pi^2}\int d^3x\; \ve_{ijk}\,{\rm Tr}\(R_i\,\(\partial_j R_k-\partial_k R_j\)\) 
 \nonumber\\
 &=& -\frac{\kappa}{24\,\pi^2}\int d^3x\; \ve_{ijk}\,R^a_i\,\partial_j R^a_k
 \equiv -\frac{\kappa}{24\,\pi^2}\int d^3x\; {\cal A}_i^a\, {\widetilde{\cal A}}_i^a
\er
where we have introduced the quantities
\be
{\cal A}_i^a\equiv R_i^b\, f_{ba} \; ; \qquad\qquad\qquad \qquad 
{\widetilde{\cal A}}_i^a\equiv f^{-1}_{ab}\,\ve_{ijk}\,\partial_j R^b_k
\ee
where $f_{ab}$ is some invertible matrix, such that $f\,f^T=h$, with $h$ being the matrix appearing in the action \rf{model}. Therefore, the static energy associated to \rf{model} can be written as
\be
E=\frac{1}{2}\,\int d^3x\; \left[ m_0^2\, h_{ab}\,R^a_{i}\,R^{b}_i+\frac{1}{e_0^2}\, h^{-1}_{ab}\,S_i^a\,S_i^b\right]
=\frac{1}{2}\,\int d^3x\; \left[ m_0^2\, \({\cal A}_i^a\)^2+\frac{1}{e_0^2}\,\({\widetilde{\cal A}}_i^a\)^2\right]
\lab{staticenergy}
\ee
where we have denoted 
\be
S_i^a=\ve_{ijk}\,\partial_jR^a_k = \frac{1}{2}\, \ve_{ijk}\,H_{jk}^a
\lab{sdef}
\ee
The fact that $Q$, given in \rf{topcharge}, is a topological charge means that it is invariant under any smooth  variation of the field configurations. Therefore, denoting by $\chi_{\alpha}$, $\alpha=1,2,3$, the three independent $SU(2)$ fields of the theory \rf{model}, one gets that the variation of $Q$ leads to the equation
\be
\partial_j\({\cal A}_i^a\,\frac{\delta {\widetilde{\cal A}}_i^a}{\delta \partial_j \chi_{\alpha}}\)
-{\cal A}_i^a\,\frac{\delta {\widetilde{\cal A}}_i^a}{\delta \chi_{\alpha}}
+\partial_j\({\widetilde{\cal A}}_i^a\,\frac{\delta {\cal A}_i^a}{\delta \partial_j \chi_{\alpha}}\)
- {\widetilde{\cal A}}_i^a\,\frac{\delta{\cal A}_i^a}{\delta \chi_{\alpha}}=0
\lab{topeq}
\ee
Note that such a relation is satisfied by any  field configuration since it is just an identity. 
On the other hand, the static Euler-Lagrange equations following from \rf{model}, or equivalently from  \rf{staticenergy}, are given by
\be
m_0^2\,\partial_j\({\cal A}_i^a\,\frac{\delta {\cal A}_i^a}{\delta \partial_j \chi_{\alpha}}\)
-m_0^2\,{\cal A}_i^a\,\frac{\delta {\cal A}_i^a}{\delta \chi_{\alpha}}
+\frac{1}{e_0^2}\,\partial_j\({\widetilde{\cal A}}_i^a\,\frac{\delta {\widetilde{\cal A}}_i^a}{\delta \partial_j \chi_{\alpha}}\)
- \frac{1}{e_0^2}\,{\widetilde{\cal A}}_i^a\,\frac{\delta {\widetilde{\cal A}}_i^a}{\delta \chi_{\alpha}}=0
\lab{eleq}
\ee
Consequently, if one imposes the fields to satisfy the following first order differential equations 
\be
{\widetilde{\cal A}}_i^a= \pm \, m_0\,e_0\,{\cal A}_i^a\; ; \qquad\qquad \rightarrow \qquad\qquad
\ve_{ijk}\,\partial_jR^a_k=\lambda\,R_i^b\,h_{ba}
\lab{bpseqs}
 \ee
it follows that \rf{bpseqs} and \rf{topeq} together, imply the Euler Lagrange equations \rf{eleq}, where we have introduced
\be
\lambda \equiv \eta\, m_0\,e_0\;;\qquad\qquad \qquad\qquad\eta\equiv \pm 1
\lab{lambdadef}
\ee
Therefore, \rf{bpseqs} are the self-duality equations for the theory \rf{model}. In addition we can write the static energy as
\be
E=\frac{1}{2}\,\int d^3x\; \left[ m_0\, {\cal A}_i^a\mp \,\frac{1}{e_0}\,{\widetilde{\cal A}}_i^a\right]^2 
\pm \frac{m_0}{e_0}\, \int d^3x\; {\cal A}_i^a\, {\widetilde{\cal A}}_i^a
\ee
and so one gets a lower bound on the energy given by 
\be
E\geq \frac{m_0}{e_0}\,\frac{24\,\pi^2}{\kappa} \,\mid Q\mid
\ee
The self-dual solutions of \rf{bpseqs} saturate such a bound. The energy for those self-dual configurations can be written as
\be
E=m_0^2\,\int d^3x\;  h_{ab}\,R^a_{i}\,R^{b}_i= \frac{1}{e_0^2}\, \int d^3x\;  h^{-1}_{ab}\,S_i^a\,S_i^b
=\frac{m_0}{e_0}\,\frac{24\,\pi^2}{\kappa} \,\mid Q\mid
\lab{selfdualenergy}
\ee
An interesting point is that, if one considers the entries of the matrix $h_{ab}$ as independent fields, then the Euler-Lagrange equations following from the energy functional \rf{staticenergy} are
\be
m_0^2 \, R_i^a\,R_i^b - \frac{1}{e_0^2}\, h^{-1}_{ca}\, h^{-1}_{bd}\, S_i^c\,S_i^d=0
\lab{habeqs}
\ee
Note that \rf{habeqs} follows as a consequence of  the self-duality equations \rf{bpseqs}, since $S_i^b\,h^{-1}_{ba}=\pm m_0\,e_0\, R_i^a$.   Therefore,  the reasoning described above, that  has led to the self-duality equations for the Euler-Lagrange equations associated to  the $SU(2)$ fields, also implies that those same first order differential equations are the self-duality equations for the Euler-Lagrange equations associated to  $h_{ab}$, considered as independent fields. 

Note that the self-duality equations \rf{bpseqs} have the same global symmetries as the theory \rf{model}. Indeed, \rf{bpseqs} are invariant under the transformations \rf{righttransf}, and also under \rf{lefttransf} if the matrix $h$ is invariant under the adjoint representation, i.e. $d\, h\, d^T=h$, otherwise the left symmetry group is broken to a given subgroup. 

It is worth mentioning that there are 9 self-duality equations \rf{bpseqs}, for 9 unknowns, namely the 3 $SU(2)$ fields and the 6 entries of the symmetric matrix $h_{ab}$. 

In the case where the matrix $h$ is the unity matrix, the self-duality equations \rf{bpseqs} reduce to 
\be
-\frac{i}{2}\,\ve_{ijk}\,\sbr{R_j}{R_k}=\pm \, m_0\,e_0\,R_i
\ee
Those are the self-duality equations for the original Skyrme model considered in \cite{ruback}, and that is known not to possess finite energy solutions. We will shown that when the matrix $h$ is non-trivial, one can have finite energy solutions of the self-duality equations \rf{bpseqs}. Self-dual  configurations can also be attained in space-times presenting compact sub-manifolds \cite{ua845,canfora}.

\section{The conformal symmetry}
\label{sec:conformal}
\setcounter{equation}{0}

We now show that the self-duality equations \rf{bpseqs} are invariant under conformal transformations in three space dimensions. We consider the $SU(2)$ field $U$ to be a scalar under space transformations, i.e.
\be
\delta x_i=\zeta_i\;;\qquad\qquad\qquad \quad \delta U=0\;;\qquad\qquad\qquad \quad
\delta\(\partial_i U\,U^{-1}\)=-\partial_i\zeta_j \;\partial_j U\,U^{-1}
\ee
From \rf{rdef2} and \rf{hdef2} one observes that $R_i^a$ and $H_{ij}^a$ contain only first derivatives of $U$, and so 
\be
\delta R_i^a= -\partial_i\zeta_j \;R_j^a\;; \qquad\qquad \qquad \qquad
\delta H_{ij}^a= -\partial_i\zeta_k \; H_{kj}^a-\partial_j\zeta_k \; H_{ik}^a
\ee
Therefore, from \rf{sdef} one has
\be
\delta S_i^a= \partial_j\zeta_i\,S^a_j-\partial_j\zeta_j\,S^a_i 
\ee
We write the self-duality equations \rf{bpseqs} as
\be
\Lambda_i^a=  S_i^a -\lambda\, R_i^b\, h_{ab}\equiv 0 
\lab{lambdadef2}
\ee 
where $\lambda$ was defined in \rf{lambdadef}, and so
\be
\delta\Lambda_i^a= \( \partial_i\zeta_j+\partial_j\zeta_i - \partial_k\zeta_k\, \delta_{ij}\)\lambda\,R_j^b\,h_{ba}-\lambda\, R_i^b\,\delta h_{ba}
\ee
If the space transformations satisfy
\be
 \partial_i\zeta_j+\partial_j\zeta_i = 2\,D\,\delta_{ij}
 \lab{conftransf}
 \ee
 for some function $D$, and if the entries of the  matrix $h_{ab}$ transform as
 \be
 \delta h_{ab}= -D\, h_{ab}
\ee
one gets that the self-duality equations \rf{bpseqs} are invariant. It follows that \rf{conftransf} are exactly the equations that define the conformal transformations in three space dimensions. Indeed, $D=0$ for translations and rotations, $D={\rm constant}$, for dilatations, and $D$ linear in $x_i$, gives the special conformal transformations \cite{babelon,shnir}. 

One can check, using the fact that $\delta h^{-1}_{ab}=D\, h^{-1}_{ab}$, that
\br
\delta \(h_{ab}\,R_i^a\, R_i^b\)&=& - 3D\,h_{ab}\,R_i^a\, R_i^b\nonumber\\ 
\delta \(h^{-1}_{ab}\,S_i^a\, S_i^b\)&=& - 3D\,h^{-1}_{ab}\,S_i^a\, S_i^b\\ 
\delta\(R_i^a\,S_i^a\)&=&- 3\,D\, R_i^a\,S_i^a\nonumber
\er
Since the volume element transforms as $\delta \(d^3x\)=3\,D\, d^3x$, it follows that the static energy \rf{staticenergy} and the topological charge \rf{topcharge} are conformally invariant.

\section{A special type of exact solution}
\label{sec:su2solutions}
\setcounter{equation}{0}

We shall now construct solutions to the self-duality equations \rf{bpseqs} where the matrix $h_{ab}$ has the form
\be
h_{ab}=f^2\,\delta_{ab}
\lab{unityh}
\ee
with $f$ a given function to be determined from \rf{bpseqs}. Note that, in such a case the matrix $h$ is invariant under the adjoint representation of $SU(2)$, i.e. $d\,h\,d^{T}=h$, and so the global transformations \rf{righttransf} and \rf{lefttransf} are symmetries of the theory \rf{model}, which can then be written as
\be
S_f=\frac{1}{\kappa}\, \int d^4x\left[ \frac{m_0^2}{2}\, f^2\,{\rm Tr}\(R_{\mu}\)^2
+\frac{1}{4\,e_0^2}\, \frac{1}{f^2}\,{\rm Tr}\(\sbr{R_{\mu}}{R_{\nu}}\)^2\right]
\lab{model2}
\ee
The self-duality equations \rf{bpseqs} becomes in this case
\be
\ve_{ijk}\,\partial_jR_k=\lambda\,R_i\,f^2=-\frac{i}{2}\,\ve_{ijk}\,\sbr{R_j}{R_k}
\lab{nicebps}
\ee
with $\lambda$ defined in \rf{lambdadef}. We now use the parameterization of the $SU(2)$ group elements given in \cite{skyrmejoaq}, i.e.
\be
U=e^{-i\,\chi_a\,\sigma_a}=e^{-i\,\chi\,T}=\one\, \cos \chi-i\, T\, \sin\chi
\lab{uparameter}
\ee
with $\sigma_a$, $a=1,2,3$, being the Pauli matrices, and 
\br
\chi=\sqrt{\chi_1^2+\chi_2^2+\chi_3^2}\;;\qquad\qquad\qquad
T=\frac{1}{1+\u2}\,\(
\begin{array}{cc}
\u2-1&-2\,i\,u\\
2\,i\,u^*&1-\u2
\end{array}\)
\er
and where we have made the stereographic projection of the unit vector $\chi_a/\chi$ on the plane and introduced the complex field $u$ as
\be
\frac{{\vec \chi}}{\chi}=\frac{1}{1+\u2}\,\(-i\(u-u^*\),u+u^*,\u2-1\)
\ee
One can show that the $SU(2)$ group elements can be further written as
\br
U=W^{\dagger}\, e^{i\,\chi\, \sigma_3}\,W\;;
\qquad\qquad\qquad 
{\rm with} \qquad\qquad 
W=\frac{1}{\sqrt{1+\u2}}\,\(
\begin{array}{cc}
1&i\,u\\
iu^*&1
\end{array}\)
\er
Therefore, one obtains that
\br
R_i=i\,\partial_iU\,U^{-1}=-V\left[\partial_i\chi\,\sigma_3+\frac{i\,2\,\sin\chi}{1+\u2}\,
\(\partial_iu\,\sigma_{+}-\partial_iu^*\,\sigma_{-}\)\right]\,V^{\dagger}
\lab{niceri}
\er
with $V= W^{\dagger}\,e^{i\,\chi\,\sigma_3}$ and $\sigma_{\pm}=\(\sigma_1\pm i\, \sigma_2\)/2$. Replacing \rf{niceri} into \rf{nicebps} one gets the following equations 
\br
\lambda\,f^2\,\partial_i \chi&=& \frac{i\,4\,\sin^2\chi}{\(1+\u2\)^2}\,\ve_{ijk}\,\partial_j u\,\partial_ku^*
\lab{nicebpschi}\\
\lambda\,f^2\,\partial_i u&=&2\, i\, \ve_{ijk}\,\partial_j \chi\,\partial_ku
\lab{nicebpsu}
\er
together with the complex conjugate of \rf{nicebpsu}. Contracting \rf{nicebpsu} with $\partial_iu$ and then with $\partial\chi_i$ one gets that the solutions of the self-duality equations \rf{nicebps} satisfy the relations 
\be
\(\partial_iu\)^2=0\;;\qquad\qquad\qquad\qquad\qquad \partial_i\chi\,\partial_iu=0
\lab{nicerelations}
\ee
which are precisely the constraints introduced in \cite{skyrmejoaq} to obtain a sub-model of the original Skyrme model with an infinite number of conserved quantities. Here, \rf{nicerelations} are not constraints but consequences of the self-duality equations. Using \rf{nicebpsu} and its complex conjugate one gets that
\be
\ve_{ijk}\,\partial_j u\,\partial_ku^*=\frac{4\,\partial_i\chi}{\lambda^2\,f^4}\, \ve_{jkl}\,\partial_j\chi\,\partial_k u\, \partial_lu^*
\ee
Replacing that into \rf{nicebpschi} one gets that
\be
f^6= \frac{i}{\lambda^3}\, \frac{16\,\sin^2\chi}{\(1+\u2\)^2}\,
\ve_{ijk}\,\partial_i\chi\,\partial_j u\, \partial_k u^*
\lab{fresult}
\ee
which is proportional to the density of the topological charge. Indeed using \rf{niceri} one gets
\be
f^6= -\frac{2\,i}{3\,\lambda^3}\,\ve_{ijk}\,{\rm Tr}\(R_i\,R_j\,R_k\)
\lab{nicef6}
\ee
Now contracting \rf{nicebpschi} with $\partial_i\chi$, and using \rf{fresult} one gets that
\be
\(\partial_i\chi\)^2=\lambda^2\, \frac{f^4}{4}
\lab{chifrel}
\ee
 a result that will be useful below.  
We now introduce the spherical-like coordinates $\(r\,,\, w_1\,,\, w_2\)$ defined by \cite{skyrmejoaq}
\be
x_1= r\, \frac{2\,w_2}{1+w_1^2+w_2^2} \; ; \qquad 
x_2= r\, \frac{2\,w_1}{1+w_1^2+w_2^2} \; ; \qquad 
x_3= r\, \frac{\(-1+w_1^2+w_2^2\)}{1+w_1^2+w_2^2} 
\ee
The metric is given by $ds^2=s_r^2\,dr^2+s_{w_1}^2\, dw_1^2+s_{w_2}^2\, dw_2^2$, with the scaling factors being
\be
s_r=1\;;\qquad\qquad\qquad s_{w_1}=s_{w_2}= \frac{2\,r}{1+w_1^2+w_2^2}
\ee
Note that since $f$ is real, the r.h.s. of \rf{fresult} has to be positive definite. Therefore, a way of  satisfying that, and also the relations \rf{nicerelations}, is to take
\be
\chi\equiv \chi\(r\)\;;\qquad\qquad u\equiv u\(w\)\;;\qquad\qquad u^*\equiv u^*\(w^*\)\; \qquad\qquad w=w_1+i\,\eta\, w_2
\lab{ansatz}
\ee
where $\eta=\pm 1$, appearing in the definition of the complex coordinate $w$, is the same sign as the one appearing in the self-duality equations \rf{bpseqs} (see \rf{lambdadef}). Indeed, using the fact that the unity vectors are ordered such ${\hat e}_r={\hat e}_{w_1}\wedge{\hat e}_{w_2}$, one gets that 
$\ve_{ijk}\,\partial_i\chi\,\partial_j u\, \partial_k u^*={\vec \nabla}\chi\cdot\({\vec \nabla}u\wedge {\vec \nabla}u^*\)=-i\,\eta\, \frac{\(1+\mid w\mid^2\)^2}{2\,r^2}\,\partial_r\chi\,\mid \partial_w u\mid^2$, and so 
\be
f^6=\frac{1}{m_0^3\,e_0^3}\, \frac{8\,\sin^2\chi}{r^2}\, \frac{\(1+\mid w\mid^2\)^2}{\(1+\u2\)^2}\,
\partial_r\chi\,\mid \partial_w u\mid^2
\lab{funnyf}
\ee
Therefore, $f^6$ is indeed positive with the choice \rf{ansatz}. But as a consequence of the ansatz \rf{ansatz} and the relation \rf{chifrel}, it follows that $f$ can depend upon $r$ only. Therefore, from  
\rf{funnyf} one must have
\be
\frac{\(1+\mid w\mid^2\)^2}{\(1+\u2\)^2}\,\mid \partial_w u\mid^2= {\rm constant}
\ee
and the  possible solution is
\be
u=w\;;\qquad \qquad\qquad u^*=w^*
\lab{usol}
\ee
and so
\be
f^6=\frac{1}{m_0^3\,e_0^3}\, \frac{8\,\sin^2\chi}{r^2}\, \partial_r\chi
\lab{chifrel2}
\ee
Therefore $\partial_r\chi$ must be positive definite. Using \rf{chifrel} and \rf{chifrel2} one gets
\be
r\,\partial_r\chi=\sin\chi
\ee
and so
\be
\chi=2\,{\rm ArcTan}\(\frac{r}{a}\)
\lab{finalsolchi}
\ee
with $a$ and arbitrary parameter of dimension of length. So, the solution can be re-scaled freely, and that  is a consequence of the conformal symmetry of the theory. Therefore, from \rf{chifrel} one gets
\be
f^2= \frac{4}{\mid m_0\,e_0\mid}\, \frac{a}{r^2+a^2}
\lab{finalf2}
\ee
Using \rf{uparameter} we can write the final solution, in polar spherical coordinates $\(r\,,\, \theta\,,\, \vp\)$, as
\br
U= \one\,\;\frac{a^2-r^2}{a^2+r^2}-i\, \frac{2\,a\,r}{a^2+r^2}\,\(
\begin{array}{cc}
\cos \theta& \sin\theta\,e^{-i\,\eta\,\vp}\\
\sin\theta\,e^{i\,\eta\,\vp} & -\cos\theta
\end{array}\)
\er 
with $\eta$ being the sign appearing in the self-duality equations \rf{bpseqs} (see \rf{lambdadef}).
Therefore
\be
U\(r=0\)= \one \;;\qquad\qquad\qquad U\(r\rightarrow \infty\)\rightarrow -\one
\ee
The topological charge \rf{topcharge} of the solution can be evaluated using \rf{nicef6} and \rf{finalf2}, to give
\be
Q=-\eta\, \frac{4}{\pi^2}\int \frac{d^3x}{a^3}\frac{1}{\(1+r^2/a^2\)^3}=-\eta
\ee
Using \rf{selfdualenergy},  \rf{niceri}, \rf{chifrel} and \rf{finalf2} one can evaluate the energy density to get 
\be
{\cal E}= m_0^2\, h_{ab}\, R_i^a\,R_i^b= m_0^2\, f^2\, \(R_i^a\)^2= \frac{192}{a^3}\,\frac{m_0}{e_0}\,  
\frac{1}{\(1+r^2/a^2\)^3}
\ee
and so the energy is 
\be
E=\int d^3x\, {\cal E}=  48\,\pi^2\, \frac{m_0}{e_0}
\ee
Therefore the energy and topological charge densities are proportional, and are spherically symmetric. They fall  as $1/r^6$ for $r\rightarrow \infty$. 
Therefore, the choice \rf{unityh} of the matrix $h_{ab}$, that does not break the global symmetries \rf{righttransf} and \rf{lefttransf}, leads to only two self-dual Skyrmion solutions of topological charges $\pm 1$.  The difference between such self-dual solutions and the Skyrmions of topological charges $\pm 1$, of the original Skyrmion model \cite{skyrme,mantonbook}, is the profile function $\chi\(r\)$, which is exact and has a slower decay at infinity.   

\section{Conclusions}
\label{sec:conclusions}
\setcounter{equation}{0}

We have proposed a new Skyrme type model that possesses, besides the usual $SU(2)$ fields, a symmetric three dimensional matrix $h_{ab}$, that governs the couplings of the components of the Maurer-Cartan form of the original Skyrme model. The introduction of that matrix is guided by some self-duality techniques of splitting the density of the barionic topological charge, that leads to an exact self-dual sector in the proposed theory. In addition, that matrix renders the self-duality equations conformally invariant in three space dimensions, and depending upon its form it may break the internal global symmetries of the original Skyrme model.  For the case where $h_{ab}$ is proportional to the identity matrix we have constructed exact self-dual Skyrmions of unity topological charges. 

The proposed model certainly opens the way to several investigations. The nature of the symmetric matrix is still not understood and deserves further studies. It could be interpreted as new independent fields, and that would require the introduction of a kinetic term for it, that our construction did not have to rely on. On the other hand, $h_{ab}$ could correspond to low energy expectation values of fields of a more fundamental theory in high energies. Investigations in that direction would certainly be important for physical application that the proposed model may have.

We have constructed self-dual solutions for the case where $h_{ab}$ is proportional to the unity matrix only. It is very important to investigate solutions for other forms of that matrix. Note that we have not made use of the conformal symmetry in the construction of an ansatz because the structure of the self-duality equations was such that it has driven us directly to a spherically symmetric  solution of unity topological charge. The use internal and conformal symmetries may be important in designing an ansatz for more general solutions. The difficulty may rely on the fact that general forms of $h_{ab}$ tend to break the left internal symmetries of the theory.  

From the  point of view of physical applications it would be interesting to break the conformal symmetry. That certainly can be attainable by a potential or even the promotion of $h_{ab}$ to physical propagating fields. On the hand, it is important to investigate the rotational modes of the self-dual solutions and their semi-classical quantization.

\vspace{4cm}

\noindent {\bf Acknowledgments:} The author is partially supported by CNPq-Brazil.

\end{document}